\documentclass[sigconf,authorversion,nonacm]{acmart}

\usepackage{enumerate}
\usepackage{hyperref}

\AtBeginDocument{%
  }

\urldef{\provoneurl}\url{http://jenkins-1.dataone.org/jenkins/view/Documentation%20Projects/job/ProvONE-Documentation-trunk/ws/provenance/ProvONE/v1/provone.html}

\makeatletter
\newcommand{\myitem}[1]{%
\item[#1]\protected@edef\@currentlabel{#1}%
}
\makeatother

\newcommand\emphAg[1]{\textsc{#1}}
\newcommand\emphAc[1]{\textsf{#1}}

\begin{document}

\title{Provenance for Lattice QCD workflows}

\author{Tanja Auge,$^1$ Gunnar Bali,$^2$ Meike Klettke,$^1$ Bertram Ludäscher,$^3$ Wolfgang S\"oldner,$^2$ Simon Weish\"aupl,$^2$ Tilo Wettig$^2$}
\email{{firstname.lastname}@ur.de}
\email{ludaesch@illinois.edu}
\affiliation{
\institution{$^1$University of Regensburg, Faculty of Computer Science and Data Science, Germany}
\institution{$^2$University of Regensburg, Department of Physics, Germany}
\institution{$^3$University of Illinois at Urbana-Champaign, School of Information Sciences, USA}
\country{}
}
  
\renewcommand{\shortauthors}{Auge et al.}

\begin{abstract}
We present a provenance model for the generic workflow of numerical Lattice Quantum Chromodynamics (QCD) calculations, which constitute an important component of particle physics research.  These calculations are carried out on the largest supercomputers worldwide with data in the multi-PetaByte range being generated and analyzed. In the Lattice QCD community, a custom metadata standard (QCDml) that includes certain provenance information already exists for one part of the workflow, the so-called \textit{generation} of configurations. 
  
In this paper, we follow the W3C PROV standard and formulate a provenance model that includes both the generation part and the so-called \textit{measurement} part of the Lattice QCD workflow. We demonstrate the applicability of this model and show how the model can be used to answer some provenance-related research questions. 
However, many important provenance questions in the Lattice QCD community require extensions of this provenance model.
To this end, we propose a multi-layered provenance approach that combines prospective and retrospective elements.
\end{abstract}


\maketitle

\section{Introduction}
Provenance generally refers to ``any information that describes the production process of an end product, which can be anything from a piece of data to a physical object'' \cite{HDL17}. This is a challenge we also face in our application, Lattice QCD workflows. Let us give a brief introduction to this application. On the fundamental level, our understanding of nature rests on the Standard Model of elementary particles and their interactions. The Standard Model is formulated in terms of quantum field theories, including \textit{Quantum Chromodynamics} (QCD). Currently many experimental and theoretical efforts are underway to search for physics beyond the Standard Model. These searches require supporting QCD calculations that must be carried out to high precision. The preferred tool for such calculations is the numerical simulation of QCD on a space-time lattice (\textit{Lattice QCD}). Lattice QCD has evolved over more than four decades and is now a mature field with many hundreds of researchers all over the world. Similar to experiments, where one first collects data and later analyzes them, the Lattice QCD programme factorizes into three parts, \textit{generation}, \textit{measurement}, and \textit{analysis} \cite{Bali:2022mlg}. In the first part, ensembles of so-called \textit{gauge-field conﬁgurations} are generated using the \textit{Markov chain Monte Carlo method} and then stored to disk. In the second part, so-called \textit{correlation functions} that are relevant for the speciﬁc physics programme are computed on these configurations, and the resulting data are also stored to disk. In the third part, the correlation functions are combined into the observables of interest. The first two parts are very compute-intensive and use the largest supercomputers worldwide. At present, typical data sizes for a given collaboration are about one PetaByte of ensembles and several PetaBytes of derived data \cite{Bali:2022mlg}. 

Already twenty years ago the Lattice QCD community initiated the \textit{International Lattice Data Grid} (ILDG) \cite{Irving:2003uk,Maynard:2004wg,Joo:2006zz,KSY22} for the purpose of sharing ensembles of configurations (according to what is nowadays known as the \textit{FAIR principles} --- FAIR stands for \textbf{F}indable, \textbf{A}ccessible, \textbf{I}nteroperable, and \textbf{R}eusable, where Reusable includes provenance aspects \cite{FAIR:2016, gofair, PUNCH4NFDI}). For the generation part of the Lattice QCD programme, the ILDG metadata standard (QCDml \cite{Maynard:2004wg,qcdml}) also includes tracking information, but a full provenance concept has not been developed yet. In this paper we present a provenance model for the generation part using the W3C PROV standard (which did not exist when the ILDG metadata standard was created). Furthermore, we extend our model to include the measurement part of the Lattice QCD programme, for which no community effort on provenance has been made so far. We view this as an important step towards implementing the FAIR principles in research data management. 

To illustrate the importance of provenance in the Lattice QCD context, let us discuss an example that occurred in practice. A set of configurations was stored at an external research institute. During the storage period, silent data corruption took place due to a file-system problem. Measurements based on the corrupted configurations could have been performed before the data corruption was noticed. In such a situation, provenance can trace incorrect measurement results back (upstream) to the corrupted configurations.  Conversely, provenance can identify the (downstream) measurement results that may have been affected by the corrupted configurations.

Another important aspect is the reproducibility and replicability (as defined in \cite{ArtifactReview}) of results published in a scientific article. On the one hand, we need provenance information to determine what part of the data was actually relevant for the published results. On the other hand, we also need provenance of the associated workflows. This will allow us (or others) to reconstruct a published result in a transparent way. For details we refer to \cite{AH19,AHH22}.

\newpage
To integrate provenance into Lattice QCD workflows we need a sustainable provenance model. After analyzing other physical problems such as \cite{SMHS22,JPDLKCS21} we decided to use the W3C PROV model \cite{PROVOverview}, which defines ``a data model, serializations, and definitions to support the interchange of provenance information on the Web.'' In \cite{JPDLKCS21}, the authors ``lay the foundation for making an automated provenance generation tool for astronomical/data-processing pipelines'' but do not use the W3C PROV model, while \cite{SMHS22} describes a W3C PROV model to make the software BACARDI, which provides a database with information about active and inactive objects orbiting the Earth, provenance-aware. The model is implemented using the Python library \texttt{prov} \cite{bacardi}. 

The provenance model towards which researchers naturally gravitate starts from a high-level conceptual workflow model and then specializes or instantiates details to answer provenance questions. Therefore, to generate our provenance model we first describe a generic Lattice QCD workflow (see Figure~\ref{fig:process} below). In a second step, we analyze a concrete Lattice QCD workflow in more detail and develop an initial W3C PROV model for it (Figure \ref{fig:example-provenance}).  Finally, we propose a multi-layered model (Figure~\ref{fig:layers}) that we envisage to be realized as a W3C PROV extension.

The structure of this paper is as follows. We first provide background information on provenance and the W3C PROV standard (Section~\ref{sec:basics}). Next, we introduce our use case (Lattice QCD) and establish our W3C PROV-based provenance model (Section~\ref{sec:Provenance}). We then discuss new ideas to extend our provenance model (Section~\ref{sec:Beyond}). Finally, we summarize and give an outlook (Section~\ref{sec:Summary}).

\paragraph{\textbf{Contributions}} Our main contributions are as follows.
\begin{itemize}
\item We demonstrate the applicability of provenance modeling for the data-intensive science field of Lattice QCD. 
\item We show how the W3C PROV model for Lattice QCD workflows can be used to answer common provenance-related research questions such as \ref{item:Q1} to \ref{item:Q5} (Section \ref{sec:Provenance}).
\item We articulate the need for model extensions to support researchers who wish to employ provenance to make their research workflows more transparent and to collect and utilize more detailed provenance information.
\end{itemize}

\section{Background and Related Work}
\label{sec:basics}
\begin{figure}[t]
\centering
\includegraphics[width=0.65\columnwidth]{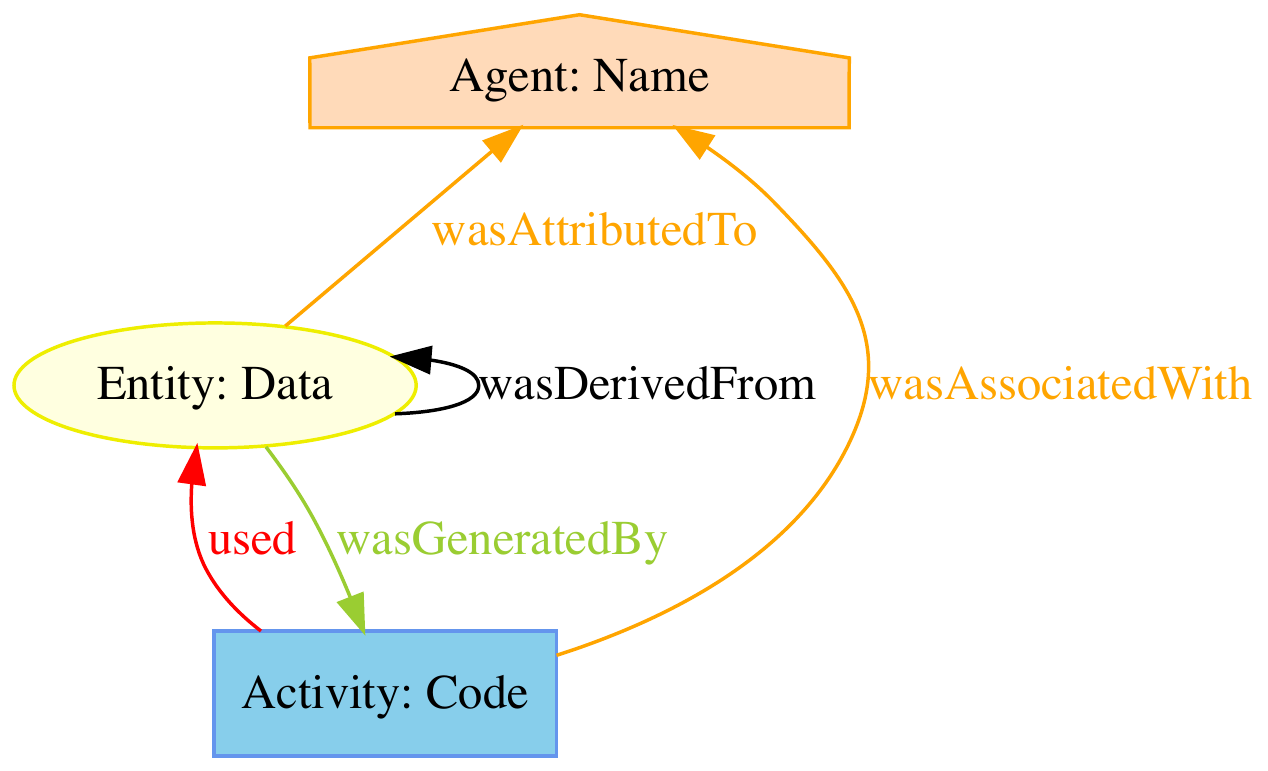}
\caption{PROV core concepts and relations, based on~\cite{PROV13}.}
\label{fig:prov-standard}
\end{figure}

Provenance information encompasses metadata on entities, activities, and agents involved in a production process. In scientific workflows this provenance information is usually logged \cite{db-spektrum-4gen22}. In the case of Lattice QCD, we are dealing with a complex workflow that generates and analyzes large amounts of data using supercomputers. For answering our provenance-related questions, which we will present in detail in Section~\ref{sec:Provenance}, we employ the W3C PROV model to describe the corresponding provenance information. This model was developed ten years ago and defines a ``core data model for provenance for building representations of the entities, people and processes involved in producing a piece of data or thing, which can be used to form assessments about its quality, reliability or trustworthiness''~\cite{PROV13}. The model distinguishes three core concepts: \textit{entities}, \textit{activities}, and \textit{agents}. Entities are data or artifacts and can be \emph{derived from} other entities. Activities can \emph{generate} or \emph{use} entities. Agents can perform or control activities or produce entities. Workflow provenance can be described by graphs whose nodes are entities, activities, or agents. Their relations are described by different  edge types of the graph (Figure \ref{fig:prov-standard}). 

At the core of the W3C PROV standard lies the  PROV data model (PROV-DM \cite{PROV-DM}). It defines concepts for expressing and exchanging provenance information and is realized by a family of related specifications, e.g.,  for provenance aimed at human consumption (PROV-N), a PROV ontology (PROV-O), and an XML schema (PROV-XML), see \cite{PROV13} for details. A direct precursor of the W3C PROV model was the Open Provenance Model (OPM \cite{MCFFGGKMMMPSSB11}).

In the context of scientific workflows, provenance is a record of the derivation of a set of results. There are two different forms of provenance \cite{ZWF06}: The W3C PROV model as well as OPM capture \textit{retrospective provenance}, i.e., information about past workflow executions and data derivations. In contrast, \textit{prospective provenance} captures the structure of a workflow and can be understood as  a recipe for future workflow executions. Workflow graphs can also serve  as a high-level  ``summary'' of what has happened in the past, i.e., despite their prospective nature, such graphs  can also be used to describe past workflow executions at a higher, conceptual level. Since the W3C PROV model (and similarly OPM) were meant to provide a minimal model for retrospective provenance, questions that involve prospective provenance or hybrid provenance elements cannot be directly answered in this model~\cite{LLCF10,SGM11}.

To address these limitations, several extensions to retrospective provenance standards have been developed over the years:  ProvONE~\cite{PROV-ONE} extends the W3C PROV model by adding a workflow (i.e., prospective provenance) layer that is then linked to the retrospective layer to support hybrid queries combining both provenance types in scientific workflow applications.  ProvONE ``aims to provide the fundamental information needed to understand and analyze scientific workflow-based computational experiments''~\cite{PROV-ONE}.
Similarly, its precursor D-PROV~\cite{MDBCL13}, and the related \emph{Open Provenance Model for Workflows} OPMW~\cite{PROV-OPMW}, provide scientists with a vocabulary and relational structure for answering hybrid provenance questions. The Wf4Ever research object model includes a vocabulary for workflow execution provenance \cite{Wf4Ever}. Numerous tools have been developed that capture computational provenance, see, e.g., \cite{pimentel_survey_2019} for a survey on provenance tools for scripts, and \cite{CSOOODM13} for a tool for capturing workflow provenance. The idea of combining retrospective with prospective provenance  to support hybrid queries has also been employed before, see,  e.g.,  \cite{zhang_revealing_2017} or \cite{PDMBKMBL16}. The latter reference combines the annotation-based prospective provenance modeling tool YesWorkflow~\cite{MPSKABB15} with a tool for capturing fine-grained retrospective provenance from Python scripts \cite{MBCKF14}.

\section{Provenance for Lattice QCD}
\label{sec:Provenance}
In this section, we develop a provenance model for Lattice QCD using the  vocabulary known in that community. A W3C PROV representation of our  model is shown in Figure \ref{fig:example-provenance} below.

\subsection{Workflow for Lattice QCD Calculations}
\label{subsec:Workflow}
As outlined in the introduction, a generic Lattice QCD workflow consists of three main parts. In the first part (\textit{generation}), so-called \textit{gauge field configurations} are generated by means of \textit{Monte Carlo} techniques, usually employing the HMC algorithm \cite{DKPR87}. At the end of this process a certain number of configurations becomes available, where each configuration consists of a fixed number of complex numbers. These configurations are stored on disk for subsequent analysis. In the second part (\textit{measurement}), \textit{correlation functions} and other derived data (collectively called \textit{measurement data}) are computed from the configurations. These correlation functions contain information about physical observables. In the third part (\textit{analysis}), the observables are computed from the correlation functions, which includes averaging over configurations, extrapolating to certain limits, and other activities. The outcome of these calculations can then be confronted with experimental results.

\begin{figure}[ht]
\centering
\includegraphics[width=0.7\columnwidth]{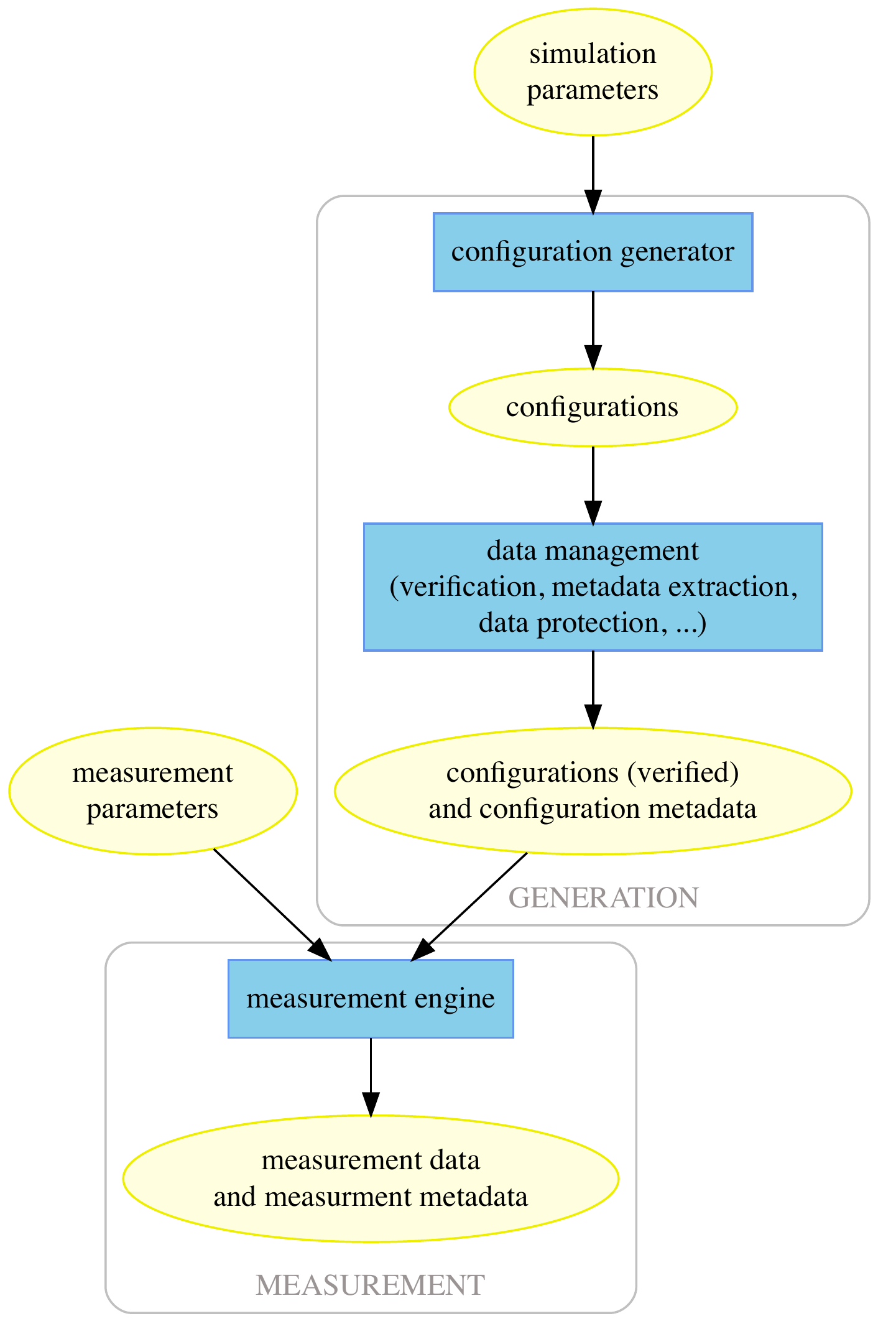}
\caption{Representation of the first two parts of a generic Lattice QCD workflow.}
\label{fig:process}
\end{figure}

The first two parts of the workflow are illustrated in Figure~\ref{fig:process} using gray frames. Here, we do not consider the third part for two reasons: the first two parts are much more compute-intensive, and the third part depends on the observable of interest and is thus much less generic than the first two. We leave the construction of a provenance model for the analysis part to future work. 

The input for the generation part is given by simulation parameters which control both the HMC algorithm (\textit{algorithmic parameters}) and the details of the physics (\textit{physical parameters}). Note that the physical parameters also enter the measurement part. Typically, they are available in the \textit{configuration metadata} that are produced during the generation part. Additional parameters (\textit{measurement parameters}) enter the measurement part. 

Let us briefly discuss the activity labeled ``data management'' in Figure~\ref{fig:process}. Typical Lattice QCD calculations generate a huge amount of data \cite{Bali:2022mlg}, and thus data management has become of critical importance \cite{KSY22}. Collecting and processing metadata is obviously a mandatory task. Furthermore, to ensure correctness, data verification is necessary, which utilizes the metadata. Also, since generating the configurations is computationally very expensive, it is important to back up the data to prevent data loss. To comply with scientific standards, often data need to be archived for a certain period of time. Note that data management is equally important for both configurations and measurement data, i.e., the measurement part of Figure~\ref{fig:process} also contains data management even though it is not shown explicitly.

\subsection{Including Provenance}
\label{sec:IncludingProv}
To better support researchers in their computational science  work, questions like the following should be answerable by the data management system and cyberinfrastructure. 
\begin{enumerate}
\myitem{\textbf{Q1}}\label{item:Q1} Which datasets are affected by an error or bug?
\myitem{\textbf{Q2}}\label{item:Q2} How are datasets affected by modifying a parameter?
\myitem{\textbf{Q3}}\label{item:Q3} Who was involved in generating the data?
\myitem{\textbf{Q4}}\label{item:Q4} Which codes and experts are needed to repeat a workflow?
\myitem{\textbf{Q5}}\label{item:Q5} Which data/parameters are needed to (re-)\,produce a result?
\end{enumerate}
These questions are just  a few examples of provenance-related  questions that  computational scientists often need to answer and are similar to those raised in \cite{SMHS22}. They are always \textit{entity-}, \textit{activity-} or \textit{agent-}focused and are used, e.g., for defect detection, quality assurance, process validation, monitoring, statistical analysis, developer evaluation or information gathering.\footnote{For other computational science provenance questions see, e.g., \cite{mcphillips_retrospective_2015} and \cite{PDMBKMBL16}.} Because of the broad range of questions, different provenance elements are required to answer the questions stated above. While questions \ref{item:Q1} and \ref{item:Q2} pertain mainly to the data level, questions \ref{item:Q3} to \ref{item:Q5} relate more closely to the underlying workflow that generated the data. 

\begin{figure}[t]
\centering
\includegraphics[width=1\columnwidth]{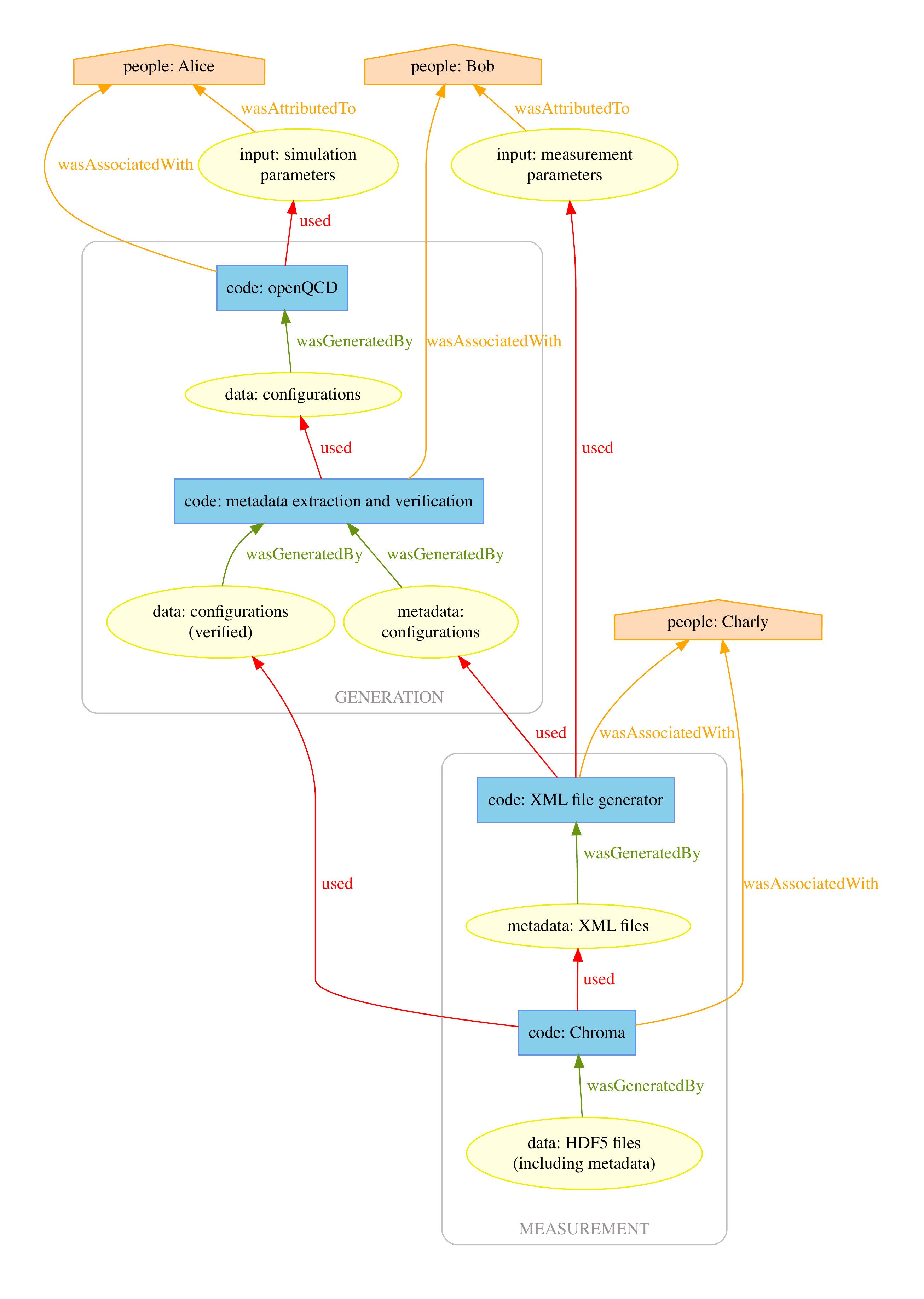}
\caption{Lattice QCD workflow as a W3C PROV model.}
\label{fig:example-provenance}
\end{figure}

We follow the W3C PROV standard presented in \cite{PROV13} to define a provenance model for the specific Lattice QCD workflow implemented in our research group, see Figure~\ref{fig:example-provenance}. The four \textit{activities} shown in the blue code boxes read the two sets of input parameters, manage the data, and generate the final HDF5 files. The seven \textit{entities} shown in the yellow boxes represent input parameters, data, and metadata. The model is completed by three \textit{agents} (orange boxes) who execute the activities and are responsible for the input parameters. In our case, the agents \emphAg{Alice} and \emphAg{Bob} define the input parameters and execute the first two activities \emphAc{openQCD} and \emphAc{metadata extraction and verification}, respectively. In the workflow, verified data and metadata are generated. Using these and the measurement parameters, \emphAg{Charly} then executes the activities in the measurement part and generates the actual HDF5 files.

There are two main differences between Figures~\ref{fig:process} and \ref{fig:example-provenance}. First, the data management activity in Figure \ref{fig:process} also includes backup and archiving, which we omitted in Figure~\ref{fig:example-provenance} in the interest of simplicity. Second, while the measurement sub-workflow in Figure~\ref{fig:process} consists of a single activity, the corresponding sub-workflow in Figure~\ref{fig:example-provenance} includes  a second activity. This is due to the fact that in the actual implementation of the measurement engine, another intermediate step is required, i.e., the generation of suitable XML input files. In summary, the provenance graph in Figure \ref{fig:example-provenance} can be seen as a refinement of the more abstract workflow version in Figure~\ref{fig:process}.

Upon closer inspection, we notice that the PROV graph does not contain  $w$-edges (\emph{wasDerivedFrom}) of the form $(D_1){\stackrel{w}\leftsquigarrow}(D_2)$, i.e., a data entity $D_2$ was derived from another entity $D_1$. Instead, Figure~\ref{fig:example-provenance} contains chains of $u$-edges (\emph{used}) and $g$-edges (\emph{wasGeneratedBy}) of the form $$(D_1)\stackrel{u}{\leftsquigarrow}[A]\stackrel{g}{\leftsquigarrow}(D_2)~,$$ i.e., a process (activity) $A$ used data entity $D_1$ and generated data entity $D_2$. In many applications, including ours, the \emph{used} and \emph{wasGeneratedBy} relations are important to explicitly model the flow of data entities in and out of processing steps (activities), thereby supporting powerful provenance analysis queries. Although the standard does not assume that a \emph{used-wasGeneratedBy} chain always implies a \emph{wasDerivedFrom} relation, this is often the case in practice and in our Lattice QCD provenance model as well. For visual clarity, we  omitted these \emph{wasDerivedFrom} edges in Figure~\ref{fig:example-provenance}. An implementation of the model could construct these edges on demand using a custom derivation rule; see  \cite{DKBL12} and \cite{moreau_rationale_2015} for further details on the interplay of these different relationships.

The provenance graph in Figure \ref{fig:example-provenance} contains specific instance-level information such as the names of the agents \emphAg{Alice}, \emphAg{Bob}, and \emphAg{Charly} and abstract identifiers on the schema level such as \texttt{XML\,file}. In many cases adding such instance-level information satisfies the needs of the Lattice QCD community. For example, the calculations are usually carried out by the same people so that the individuals/concrete agents can be viewed as ``part of the system.'' 

Figure \ref{fig:example-provenance} lists the names of the individuals who regularly act as agents for a specific research group and time period. A more generic setup would indicate an agent of type \texttt{Person}. Using the names of individuals, question \ref{item:Q3} can be answered directly from the provenance graph. In contrast, question \ref{item:Q4} refers to activities and agents at the workflow level and requires prospective provenance. \ref{item:Q5} is a similar question posed at the entity level.

Like \ref{item:Q3}, question \ref{item:Q1} is a retrospective provenance question. For example, we may realize that the output data are incorrect. In this case, the data derivation chains must be traced from the results back to the sources (\emph{upstream} propagation). As another example, we may find an error in an activity or entity, such as the silent data corruption in the configurations mentioned above. Then the erroneous activity or entity needs to be fixed, and subsequent entities need to be recomputed or corrected (\emph{downstream} propagation). In both examples we have to perform dependency  tracing along derivation chains in the provenance graph.

Question \ref{item:Q2} can be interpreted in different ways. If we are solely interested in the dependency structure at the conceptual level (Figure \ref{fig:example-provenance}), question \ref{item:Q2} requires prospective provenance only. Alternatively, if we are interested in the effect of a parameter change on a dataset in a previous workflow run, we need retrospective provenance. 

To summarize, questions \ref{item:Q1} and \ref{item:Q3} require retrospective provenance, questions \ref{item:Q4} and \ref{item:Q5} correspond to prospective provenance queries, and question \ref{item:Q2} combines both kinds of provenance.

\section{Towards Layered Provenance}
\label{sec:Beyond}

Every execution of the workflow by \emphAg{Alice}, \emphAg{Bob}, and \emphAg{Charly}  results in a provenance graph similar to the one shown in Figure~\ref{fig:example-provenance}. Since the overall provenance model structure remains the same in all cases, the provenance graph depicted in Figure~\ref{fig:example-provenance} is really a \emph{provenance template} graph, i.e., each \emph{workflow run} (execution) generates its own \emph{provenance instance} graph, in which schema-level elements (e.g., \texttt{data:HDF5\,files}) are replaced by references to concrete instance objects (e.g., \texttt{X251r000n1000\_run3.hd5}). In turn, the provenance template in Figure~\ref{fig:example-provenance} can be seen as a specialization  of the workflow graph in Figure~\ref{fig:process} that describes the general form of Lattice QCD workflows used by the community. For example, the generic steps \emph{configuration\,generator} and \emph{measurement engine} in the workflow are specialized to \emph{code:\,openQCD} \cite{Luscher:2012av} and \emph{code:\,Chroma} \cite{Edwards:2004sx}, respectively, which are the specific tools used by the physicists in our research group.
 
\subsection{Provenance Templates vs Instances}
To address the practical needs of our research scientists, while at the same time employing a standard  model to facilitate data exchange and transparency, we propose to extend the W3C PROV model to include both instance-level provenance graphs and---linked to these---a template-level provenance graph. The relationship between template and instance graphs is a very natural one, as the latter can be viewed as isomorphic copies of the former,  where schema-level elements have been replaced  by object identifiers. In this way, the template graph can serve as an overview or a summary of the many instance graphs. 

\begin{figure}[t]
\centering
\includegraphics[width=\columnwidth]{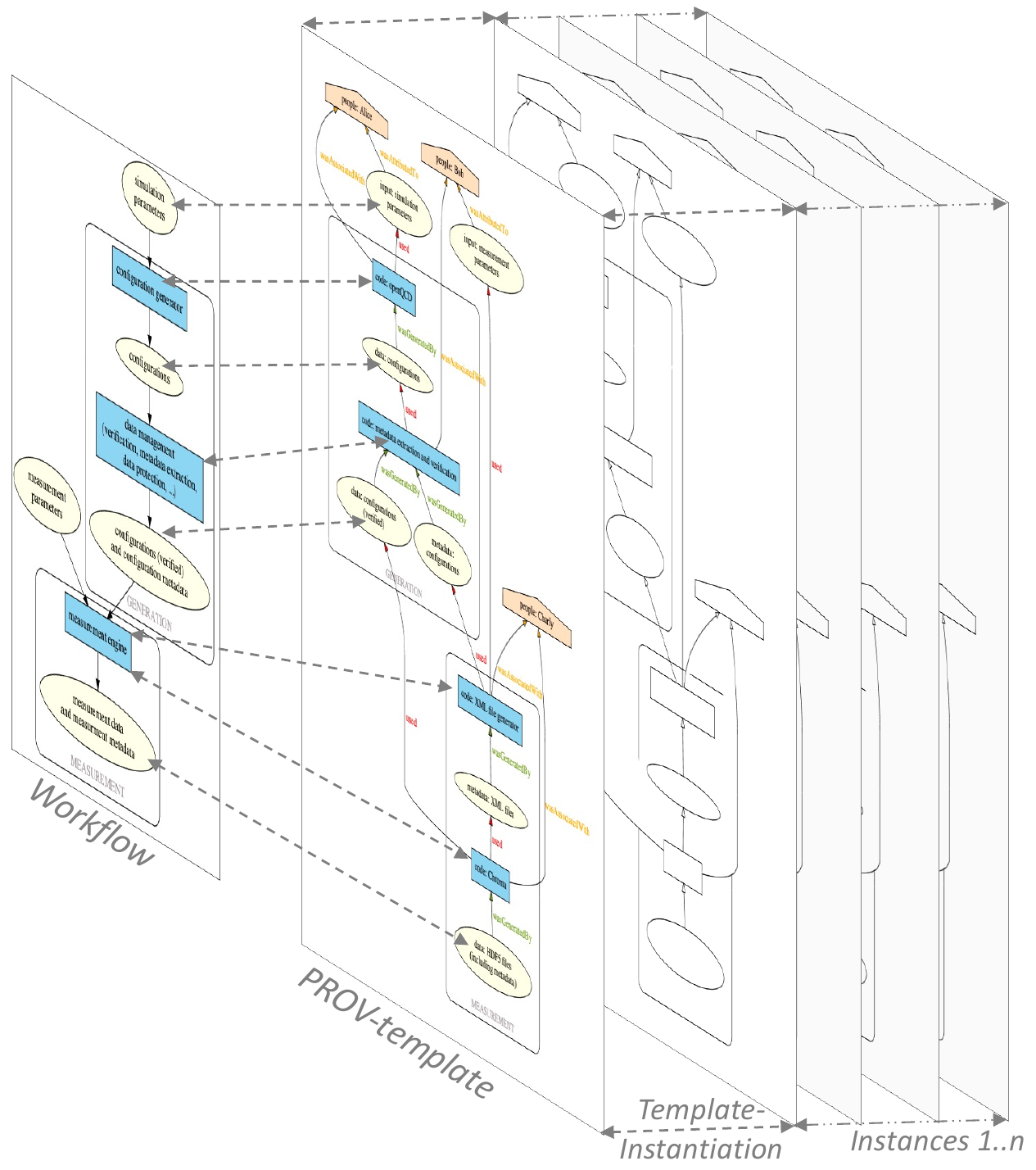}
\caption{\boldmath Layers of the proposed hybrid provenance model:  Elements of a conceptual workflow  (left) naturally map to  elements in a W3C PROV template model (middle). Each workflow execution creates PROV-compatible retrospective provenance  graphs (right), i.e., instances $1,\dots, n$ of the provenance template in the middle. }
\label{fig:layers}
\end{figure}

We further propose to add a workflow layer to this extension. In the resulting multi-layer provenance model (Figure~\ref{fig:layers}), a community-wide \emph{workflow graph} can be specialized to a \emph{provenance template} (for individual research groups), which in turn will be \emph{instantiated} whenever workflow runs are executed. These instances then include concrete values of all input parameters, the names and time stamps of the data files containing the configurations, the version numbers or git hashes of the codes, compiler versions and flags, the names of persons who executed the compilation, details of the machines running the calculations, etc. 

As our initial analysis of the provenance-related queries \ref{item:Q1} to \ref{item:Q5} from Section~\ref{sec:IncludingProv} has shown,  to answer all such questions requires a combination of retrospective provenance elements (as provided by the W3C PROV standard) and prospective elements (as given by a workflow or a provenance template graph). After extending our  model in this way,  it should satisfy the following desiderata:
\begin{enumerate}
\item The community-level workflow  structure should be linkable to provenance template graphs (research group level). 
\item The instance-level provenance graphs of the hundreds or thousands of runs (with varying parameter settings etc.) should be automatically linked to a provenance template.
\item A domain-aware provenance model should allow users to distinguish different types of data, e.g.,  using namespaces.
\end{enumerate}

\noindent The proposed multi-layer model (Figure~\ref{fig:layers}) can be implemented in different ways. We hope to bring together both communities, i.e., our colleagues from the Lattice QCD community and the provenance research community, to co-develop suitable W3C PROV extensions. Our current plan is to explore and evaluate existing standards, in particular W3C PROV and RDF \cite{RDF}. The latter would also allow us to embed multi-layered provenance models into a more general \emph{knowledge graph}/LOD (Linked Open Data) framework, leveraging again existing standards, tools, and namespaces.
 Using these, the Lattice QCD community can determine how far they want to go with modeling and formalizing the semantics for the different types, e.g., to create controlled vocabularies and/or  formal ontologies in OWL, agreed upon by the community. We reach out to the provenance community to get a head-start in our model-building efforts. Since we are not the first ones to identify the need for \emph{hybrid provenance} models (see, e.g., \cite{ZWF06,PROV-ONE,PROV-OPMW,CSOOODM13,Wf4Ever} among many others), we hope to build on existing efforts rather than reinventing the wheel.

\subsection{Implementation \& Evaluation Plan}
In order to evaluate the efficacy and practicality of the proposed model, we plan to implement a prototype for the Lattice QCD community. Since a considerable amount of provenance information is already captured by existing QCD workflows in log files, we will first develop a Python-based provenance harvesting tool. The harvested information then needs to be mapped to a suitable provenance store (e.g., a relational or graph database) that implements our model. Additional provenance information can be found in the attributes and dataspace objects of the HDF5 datasets and in the file and folder names. As described in \cite{mcphillips_retrospective_2015}, this provenance information will then be extracted based on the applicable conventions. Finally, provenance information that is required in our model, but not (yet) available through harvesting, will have to be recorded through other means, e.g., using a light-weight provenance recorder, through instrumentation of the code, or by writing additional information to log files.  

Our development efforts will be informed by a survey on collecting and managing provenance from scripts \cite{pimentel_survey_2019} and practical experience with tools that capture and integrate prospective and retrospective provenance information \cite{MPSKABB15,MBCKF14,PDMBKMBL16,zhang_revealing_2017}. For exporting interoperable provenance in W3C PROV-compliant form we will use the Python \texttt{prov} library \cite{prov-library}. We will also explore the option to cast our extended provenance model as a W3C standards-compliant knowledge graph. This would allow us to leverage additional standards and tools based on popular triple stores and graph query languages such as SPARQL \cite{PROV-sparql} or Cypher \cite{opencypher}.

\section{Summary and Future Work}
\label{sec:Summary}
Lattice QCD is an important field of particle physics that generates and analyzes huge amounts of data. We have proposed a provenance model for Lattice QCD workflows based on the W3C PROV standard. Starting from a generic workflow (Figure~\ref{fig:process}) we have derived a PROV template (Figure~\ref{fig:example-provenance}) that allows us to answer a number of typical provenance-related questions. To address a more complete set of provenance questions (see, e.g.,  \ref{item:Q1} to \ref{item:Q5}) we have proposed a layered model that provides the necessary information. It consists of a workflow layer, a provenance template layer, and an instance layer (Figure~\ref{fig:layers}).

In future work, we plan to apply our provenance model to the third part (\emph{analysis}) of the overall Lattice QCD workflow. As indicated earlier, this part is less generic and more tailored to the specific physics programme of a particular collaboration. 

We have argued in Section~\ref{subsec:Workflow} that data management is important in all parts of the Lattice QCD workflow. Therefore, it is desirable to build upon the Lattice QCD community efforts in ILDG and PUNCH4NFDI to define more comprehensive standards for data management, which also need to include provenance metadata.

 We plan to engage the Lattice QCD and provenance communities to refine our initial model proposal and, subsequently, to implement a prototype that will allow us to evaluate its efficacy and practicality.

\section*{Acknowledgments} We thank Sara Collins, Christoph Lehner, Nils Meyer and Stefan Solbrig for stimulating discussions. This work was supported in part by the Deutsche Forschungsgemeinschaft (DFG) under grant
``NFDI 39/1'' (PUNCH4NFDI).

\bibliographystyle{ACM-Reference-Format}
\bibliography{literatur}


\begin{thebibliography}{44}


\ifx \showCODEN    \undefined \def \showCODEN     #1{\unskip}     \fi
\ifx \showDOI      \undefined \def \showDOI       #1{#1}\fi
\ifx \showISBNx    \undefined \def \showISBNx     #1{\unskip}     \fi
\ifx \showISBNxiii \undefined \def \showISBNxiii  #1{\unskip}     \fi
\ifx \showISSN     \undefined \def \showISSN      #1{\unskip}     \fi
\ifx \showLCCN     \undefined \def \showLCCN      #1{\unskip}     \fi
\ifx \shownote     \undefined \def \shownote      #1{#1}          \fi
\ifx \showarticletitle \undefined \def \showarticletitle #1{#1}   \fi
\ifx \showURL      \undefined \def \showURL       {\relax}        \fi
\providecommand\bibfield[2]{#2}
\providecommand\bibinfo[2]{#2}
\providecommand\natexlab[1]{#1}
\providecommand\showeprint[2][]{arXiv:#2}

\bibitem[ACM(2020)]%
        {ArtifactReview}
\bibfield{author}{\bibinfo{person}{ACM}.} \bibinfo{year}{2020}\natexlab{}.
\newblock \bibinfo{title}{Artifact Review and Badging - Current}.
\newblock
\newblock
\urldef\tempurl%
\url{https://www.acm.org/publications/policies/artifact-review-and-badging-current}
\showURL{%
\tempurl}


\bibitem[Auge et~al\mbox{.}(2022)]%
        {AHH22}
\bibfield{author}{\bibinfo{person}{Tanja Auge}, \bibinfo{person}{Moritz
  Hanzig}, {and} \bibinfo{person}{Andreas Heuer}.}
  \bibinfo{year}{2022}\natexlab{}.
\newblock \showarticletitle{{ProSA Pipeline: Provenance Conquers the Chase}}.
  In \bibinfo{booktitle}{\emph{{ADBIS} (Short Papers)}}
  \emph{(\bibinfo{series}{CCIS}, Vol.~\bibinfo{volume}{1652})}.
  \bibinfo{publisher}{Springer}, \bibinfo{pages}{89--98}.
\newblock


\bibitem[Auge and Heuer(2019)]%
        {AH19}
\bibfield{author}{\bibinfo{person}{Tanja Auge} {and} \bibinfo{person}{Andreas
  Heuer}.} \bibinfo{year}{2019}\natexlab{}.
\newblock \showarticletitle{{ProSA - Using the CHASE for Provenance
  Management}}. In \bibinfo{booktitle}{\emph{{ADBIS}}}
  \emph{(\bibinfo{series}{LNCS}, Vol.~\bibinfo{volume}{11695})}.
  \bibinfo{publisher}{Springer}, \bibinfo{pages}{357--372}.
\newblock


\bibitem[Bali et~al\mbox{.}(2022)]%
        {Bali:2022mlg}
\bibfield{author}{\bibinfo{person}{Gunnar Bali} {et~al\mbox{.}}}
  \bibinfo{year}{2022}\natexlab{}.
\newblock \showarticletitle{{Lattice gauge ensembles and data management}}.
\newblock \bibinfo{journal}{\emph{Proceedings of Science (PoS)}}
  \bibinfo{volume}{LATTICE} (\bibinfo{year}{2022}), \bibinfo{pages}{203}.
\newblock
\showeprint[arxiv]{2212.10138}


\bibitem[Costa et~al\mbox{.}(2013)]%
        {CSOOODM13}
\bibfield{author}{\bibinfo{person}{Flavio Costa} {et~al\mbox{.}}}
  \bibinfo{year}{2013}\natexlab{}.
\newblock \showarticletitle{Capturing and querying workflow runtime provenance
  with {PROV:} a practical approach}. In \bibinfo{booktitle}{\emph{{EDBT/ICDT}
  Workshops}}. \bibinfo{publisher}{{ACM}}, \bibinfo{pages}{282--289}.
\newblock


\bibitem[Cuevas-Vicenttín et~al\mbox{.}(2016)]%
        {PROV-ONE}
\bibfield{author}{\bibinfo{person}{Víctor Cuevas-Vicenttín} {et~al\mbox{.}}}
  \bibinfo{year}{2016}\natexlab{}.
\newblock \bibinfo{title}{{ProvONE: A PROV Extension Data Model for Scientific
  Workflow Provenance}}.
\newblock \bibinfo{howpublished}{\provoneurl}.
\newblock


\bibitem[Cyganiak et~al\mbox{.}(2014)]%
        {RDF}
\bibfield{author}{\bibinfo{person}{Richard Cyganiak}, \bibinfo{person}{David
  Wood}, {and} \bibinfo{person}{Markus Lanthaler}.}
  \bibinfo{year}{2014}\natexlab{}.
\newblock \bibinfo{title}{RDF 1.1 Concepts and Abstract Syntax}.
\newblock
\newblock
\urldef\tempurl%
\url{https://www.w3.org/TR/rdf11-concepts/}
\showURL{%
\tempurl}


\bibitem[Dey et~al\mbox{.}(2012)]%
        {DKBL12}
\bibfield{author}{\bibinfo{person}{Saumen~C. Dey} {et~al\mbox{.}}}
  \bibinfo{year}{2012}\natexlab{}.
\newblock \showarticletitle{{Datalog as a Lingua Franca for Provenance Querying
  and Reasoning}}. In \bibinfo{booktitle}{\emph{TaPP}}.
  \bibinfo{publisher}{{USENIX} Association}.
\newblock


\bibitem[DLR(2022)]%
        {bacardi}
\bibfield{author}{\bibinfo{person}{DLR}.} \bibinfo{year}{2022}\natexlab{}.
\newblock \bibinfo{booktitle}{\emph{Integrating Provenance-Awareness into the
  Space Debris Processing System BACARDI}}.
\newblock
\urldef\tempurl%
\url{https://github.com/DLR-SC/bacardi-prov-model}
\showURL{%
\tempurl}


\bibitem[Duane et~al\mbox{.}(1987)]%
        {DKPR87}
\bibfield{author}{\bibinfo{person}{Simon Duane} {et~al\mbox{.}}}
  \bibinfo{year}{1987}\natexlab{}.
\newblock \showarticletitle{{Hybrid Monte Carlo}}.
\newblock \bibinfo{journal}{\emph{Phys. Lett. B}}  \bibinfo{volume}{195}
  (\bibinfo{year}{1987}), \bibinfo{pages}{216--222}.
\newblock


\bibitem[Edwards and Joó(2005)]%
        {Edwards:2004sx}
\bibfield{author}{\bibinfo{person}{Robert~G. Edwards} {and}
  \bibinfo{person}{Bálint Joó}.} \bibinfo{year}{2005}\natexlab{}.
\newblock \showarticletitle{{The Chroma software system for lattice QCD}}.
\newblock \bibinfo{journal}{\emph{Nucl. Phys. B Proc. Suppl.}}
  \bibinfo{volume}{140} (\bibinfo{year}{2005}), \bibinfo{pages}{832}.
\newblock
\showeprint[arxiv]{hep-lat/0409003}


\bibitem[Garijo et~al\mbox{.}(2014)]%
        {PROV-OPMW}
\bibfield{author}{\bibinfo{person}{Daniel Garijo}, \bibinfo{person}{Yolanda
  Gil}, {and} \bibinfo{person}{Oscar Corcho}.} \bibinfo{year}{2014}\natexlab{}.
\newblock \bibinfo{title}{OPMW-PROV: The Open Provenance Model for Workflows}.
\newblock
\newblock
\urldef\tempurl%
\url{https://www.opmw.org}
\showURL{%
\tempurl}


\bibitem[Gil et~al\mbox{.}(2013)]%
        {PROV13}
\bibfield{author}{\bibinfo{person}{Yolanda Gil} {et~al\mbox{.}}}
  \bibinfo{year}{2013}\natexlab{}.
\newblock \bibinfo{title}{{PROV Model Primer}}.
\newblock
\newblock
\urldef\tempurl%
\url{https://www.w3.org/TR/prov-primer/}
\showURL{%
\tempurl}


\bibitem[GoFair(2023)]%
        {gofair}
\bibfield{author}{\bibinfo{person}{GoFair}.} \bibinfo{year}{2023}\natexlab{}.
\newblock \bibinfo{title}{FAIR Principles}.
\newblock
\newblock
\urldef\tempurl%
\url{https://www.go-fair.org}
\showURL{%
\tempurl}


\bibitem[Groth and Moreau(2013)]%
        {PROVOverview}
\bibfield{author}{\bibinfo{person}{Paul Groth} {and} \bibinfo{person}{Luc
  Moreau}.} \bibinfo{year}{2013}\natexlab{}.
\newblock \bibinfo{title}{PROV-Overview}.
\newblock
\newblock
\urldef\tempurl%
\url{https://www.w3.org/TR/2013/NOTE-prov-overview-20130430/}
\showURL{%
\tempurl}


\bibitem[Harris et~al\mbox{.}(2013)]%
        {PROV-sparql}
\bibfield{author}{\bibinfo{person}{Steve Harris}, \bibinfo{person}{Andy
  Seaborne}, {and} \bibinfo{person}{Eric Prud'hommeaux}.}
  \bibinfo{year}{2013}\natexlab{}.
\newblock \bibinfo{title}{SPARQL 1.1 Query Language}.
\newblock
\newblock
\urldef\tempurl%
\url{https://www.w3.org/TR/sparql11-query/}
\showURL{%
\tempurl}


\bibitem[Herschel et~al\mbox{.}(2017)]%
        {HDL17}
\bibfield{author}{\bibinfo{person}{Melanie Herschel}, \bibinfo{person}{Ralf
  Diestelk{\"{a}}mper}, {and} \bibinfo{person}{Houssem {Ben Lahmar}}.}
  \bibinfo{year}{2017}\natexlab{}.
\newblock \showarticletitle{A survey on provenance: What for? What form? What
  from?}
\newblock \bibinfo{journal}{\emph{{VLDB} J.}} \bibinfo{volume}{26},
  \bibinfo{number}{6} (\bibinfo{year}{2017}), \bibinfo{pages}{881--906}.
\newblock


\bibitem[Huynh(2020)]%
        {prov-library}
\bibfield{author}{\bibinfo{person}{Trung~Dong Huynh}.}
  \bibinfo{year}{2020}\natexlab{}.
\newblock \bibinfo{title}{prov 2.0.0}.
\newblock
\newblock
\urldef\tempurl%
\url{https://pypi.org/project/prov/}
\showURL{%
\tempurl}


\bibitem[Inc.(2021)]%
        {opencypher}
\bibfield{author}{\bibinfo{person}{Neo4j Inc.}}
  \bibinfo{year}{2021}\natexlab{}.
\newblock \bibinfo{title}{What is openCypher?}
\newblock
\newblock
\urldef\tempurl%
\url{https://opencypher.org}
\showURL{%
\tempurl}


\bibitem[Irving et~al\mbox{.}(2004)]%
        {Irving:2003uk}
\bibfield{author}{\bibinfo{person}{A.C. Irving} {et~al\mbox{.}}}
  \bibinfo{year}{2004}\natexlab{}.
\newblock \showarticletitle{{Progress in building an International Lattice Data
  Grid}}.
\newblock \bibinfo{journal}{\emph{Nucl. Phys. B Proc. Suppl.}}
  \bibinfo{volume}{129} (\bibinfo{year}{2004}), \bibinfo{pages}{159--163}.
\newblock
\showeprint[arxiv]{hep-lat/0309029}


\bibitem[Johnson et~al\mbox{.}(2021)]%
        {JPDLKCS21}
\bibfield{author}{\bibinfo{person}{Michael A.~C. Johnson} {et~al\mbox{.}}}
  \bibinfo{year}{2021}\natexlab{}.
\newblock \showarticletitle{Astronomical Pipeline Provenance: {A} Use Case
  Evaluation}. In \bibinfo{booktitle}{\emph{TaPP}}.
  \bibinfo{publisher}{{USENIX} Association}.
\newblock


\bibitem[Joó and Maynard(2006)]%
        {Joo:2006zz}
\bibfield{author}{\bibinfo{person}{Bálint Joó} {and} \bibinfo{person}{C.~M.
  Maynard}.} \bibinfo{year}{2006}\natexlab{}.
\newblock \showarticletitle{{Progress in building the International Lattice
  Data Grid}}.
\newblock \bibinfo{journal}{\emph{PoS}}  \bibinfo{volume}{LAT2006}
  (\bibinfo{year}{2006}), \bibinfo{pages}{028}.
\newblock
\urldef\tempurl%
\url{https://doi.org/10.22323/1.032.0028}
\showDOI{\tempurl}


\bibitem[Karsch et~al\mbox{.}(2022)]%
        {KSY22}
\bibfield{author}{\bibinfo{person}{Frithjof Karsch}, \bibinfo{person}{Hubert
  Simma}, {and} \bibinfo{person}{Tomoteru Yoshie}.}
  \bibinfo{year}{2022}\natexlab{}.
\newblock \showarticletitle{{The International Lattice Data Grid - towards FAIR
  Data}}. In \bibinfo{booktitle}{\emph{39th Intl.\ Symposium on Lattice Field
  Theory}}.
\newblock
\showeprint[arxiv]{2212.08392}~[hep-lat]


\bibitem[Klettke and St{\"{o}}rl(2022)]%
        {db-spektrum-4gen22}
\bibfield{author}{\bibinfo{person}{Meike Klettke} {and} \bibinfo{person}{Uta
  St{\"{o}}rl}.} \bibinfo{year}{2022}\natexlab{}.
\newblock \showarticletitle{Four Generations in Data Engineering for Data
  Science}.
\newblock \bibinfo{journal}{\emph{Datenbank-Spektrum}} \bibinfo{volume}{22},
  \bibinfo{number}{1} (\bibinfo{year}{2022}), \bibinfo{pages}{59--66}.
\newblock


\bibitem[Lim et~al\mbox{.}(2010)]%
        {LLCF10}
\bibfield{author}{\bibinfo{person}{Chunhyeok Lim} {et~al\mbox{.}}}
  \bibinfo{year}{2010}\natexlab{}.
\newblock \showarticletitle{{Prospective and Retrospective Provenance
  Collection in Scientific Workflow Environments}}. In
  \bibinfo{booktitle}{\emph{{SCC}}}. \bibinfo{publisher}{{IEEE} Computer
  Society}, \bibinfo{pages}{449--456}.
\newblock


\bibitem[Lüscher and Schaefer(2013)]%
        {Luscher:2012av}
\bibfield{author}{\bibinfo{person}{Martin Lüscher} {and}
  \bibinfo{person}{Stefan Schaefer}.} \bibinfo{year}{2013}\natexlab{}.
\newblock \showarticletitle{{Lattice QCD with open boundary conditions and
  twisted-mass reweighting}}.
\newblock \bibinfo{journal}{\emph{Comput. Phys. Commun.}}
  \bibinfo{volume}{184} (\bibinfo{year}{2013}), \bibinfo{pages}{519--528}.
\newblock
\urldef\tempurl%
\url{https://doi.org/10.1016/j.cpc.2012.10.003}
\showDOI{\tempurl}
\showeprint[arxiv]{1206.2809}~[hep-lat]


\bibitem[Maynard and Pleiter(2005)]%
        {Maynard:2004wg}
\bibfield{author}{\bibinfo{person}{C.~M. Maynard} {and} \bibinfo{person}{D.
  Pleiter}.} \bibinfo{year}{2005}\natexlab{}.
\newblock \showarticletitle{{QCDml: First milestone for building an
  International Lattice Data Grid}}.
\newblock \bibinfo{journal}{\emph{Nucl. Phys. B Proc. Suppl.}}
  \bibinfo{volume}{140} (\bibinfo{year}{2005}), \bibinfo{pages}{213--221}.
\newblock
\urldef\tempurl%
\url{https://doi.org/10.1016/j.nuclphysbps.2004.11.116}
\showDOI{\tempurl}
\showeprint[arxiv]{hep-lat/0409055}


\bibitem[McPhillips et~al\mbox{.}(2015a)]%
        {MPSKABB15}
\bibfield{author}{\bibinfo{person}{Timothy McPhillips} {et~al\mbox{.}}}
  \bibinfo{year}{2015}\natexlab{a}.
\newblock \showarticletitle{{{YesWorkflow}}: {{A User-Oriented}},
  {{Language-Independent Tool}} for {{Recovering Workflow Information}} from
  {{Scripts}}}.
\newblock \bibinfo{journal}{\emph{Intl.\ Journal of Digital Curation}}
  \bibinfo{volume}{10}, \bibinfo{number}{1} (\bibinfo{year}{2015}),
  \bibinfo{pages}{298--313}.
\newblock
\urldef\tempurl%
\url{https://doi.org/10.2218/ijdc.v10i1.370}
\showDOI{\tempurl}


\bibitem[McPhillips et~al\mbox{.}(2015b)]%
        {mcphillips_retrospective_2015}
\bibfield{author}{\bibinfo{person}{Timothy~M. McPhillips} {et~al\mbox{.}}}
  \bibinfo{year}{2015}\natexlab{b}.
\newblock \showarticletitle{Retrospective Provenance Without a Runtime
  Provenance Recorder}. In \bibinfo{booktitle}{\emph{TaPP}}.
  \bibinfo{publisher}{{USENIX} Association}.
\newblock


\bibitem[Missier et~al\mbox{.}(2013)]%
        {MDBCL13}
\bibfield{author}{\bibinfo{person}{Paolo Missier} {et~al\mbox{.}}}
  \bibinfo{year}{2013}\natexlab{}.
\newblock \showarticletitle{{D-PROV:} Extending the {PROV} Provenance Model
  with Workflow Structure}. In \bibinfo{booktitle}{\emph{TaPP}}.
  \bibinfo{publisher}{{USENIX} Association}.
\newblock


\bibitem[Moreau et~al\mbox{.}(2011)]%
        {MCFFGGKMMMPSSB11}
\bibfield{author}{\bibinfo{person}{Luc Moreau} {et~al\mbox{.}}}
  \bibinfo{year}{2011}\natexlab{}.
\newblock \showarticletitle{The Open Provenance Model core specification
  (v1.1)}.
\newblock \bibinfo{journal}{\emph{Future Gener. Comput. Syst.}}
  \bibinfo{volume}{27}, \bibinfo{number}{6} (\bibinfo{year}{2011}),
  \bibinfo{pages}{743--756}.
\newblock


\bibitem[Moreau et~al\mbox{.}(2013)]%
        {PROV-DM}
\bibfield{author}{\bibinfo{person}{Luc Moreau} {et~al\mbox{.}}}
  \bibinfo{year}{2013}\natexlab{}.
\newblock \bibinfo{title}{{PROV-DM: The PROV Data Model}}.
\newblock
\newblock
\urldef\tempurl%
\url{https://www.w3.org/TR/2013/REC-prov-dm-20130430/}
\showURL{%
\tempurl}


\bibitem[Moreau et~al\mbox{.}(2015)]%
        {moreau_rationale_2015}
\bibfield{author}{\bibinfo{person}{Luc Moreau} {et~al\mbox{.}}}
  \bibinfo{year}{2015}\natexlab{}.
\newblock \showarticletitle{The rationale of {PROV}}.
\newblock \bibinfo{journal}{\emph{J. Web Semant.}}  \bibinfo{volume}{35}
  (\bibinfo{year}{2015}), \bibinfo{pages}{235--257}.
\newblock


\bibitem[Murta et~al\mbox{.}(2014)]%
        {MBCKF14}
\bibfield{author}{\bibinfo{person}{Leonardo Murta} {et~al\mbox{.}}}
  \bibinfo{year}{2014}\natexlab{}.
\newblock \showarticletitle{{noWorkflow: Capturing and Analyzing Provenance of
  Scripts}}. In \bibinfo{booktitle}{\emph{{IPAW}}}
  \emph{(\bibinfo{series}{LNCS}, Vol.~\bibinfo{volume}{8628})}.
  \bibinfo{publisher}{Springer}, \bibinfo{pages}{71--83}.
\newblock


\bibitem[Pimentel et~al\mbox{.}(2016)]%
        {PDMBKMBL16}
\bibfield{author}{\bibinfo{person}{Jo{\~{a}}o~Felipe Pimentel} {et~al\mbox{.}}}
  \bibinfo{year}{2016}\natexlab{}.
\newblock \showarticletitle{Yin {\&} Yang: Demonstrating Complementary
  Provenance from noWorkflow {\&} YesWorkflow}. In
  \bibinfo{booktitle}{\emph{{IPAW}}} \emph{(\bibinfo{series}{LNCS},
  Vol.~\bibinfo{volume}{9672})}. \bibinfo{publisher}{Springer},
  \bibinfo{pages}{161--165}.
\newblock


\bibitem[Pimentel et~al\mbox{.}(2019)]%
        {pimentel_survey_2019}
\bibfield{author}{\bibinfo{person}{Jo{\~{a}}o~Felipe Pimentel} {et~al\mbox{.}}}
  \bibinfo{year}{2019}\natexlab{}.
\newblock \showarticletitle{A Survey on Collecting, Managing, and Analyzing
  Provenance from Scripts}.
\newblock \bibinfo{journal}{\emph{{ACM} Comput. Surv.}} \bibinfo{volume}{52},
  \bibinfo{number}{3} (\bibinfo{year}{2019}), \bibinfo{pages}{47:1--47:38}.
\newblock


\bibitem[PUNCH4NFDI(2023)]%
        {PUNCH4NFDI}
\bibfield{author}{\bibinfo{person}{PUNCH4NFDI}.}
  \bibinfo{year}{2023}\natexlab{}.
\newblock \bibinfo{title}{A consortium in the {NFDI} (National Research Data
  Infrastructure)}.
\newblock
\newblock
\urldef\tempurl%
\url{https://www.punch4nfdi.de}
\showURL{%
\tempurl}


\bibitem[Simmhan et~al\mbox{.}(2011)]%
        {SGM11}
\bibfield{author}{\bibinfo{person}{Yogesh Simmhan}, \bibinfo{person}{Paul
  Groth}, {and} \bibinfo{person}{Luc Moreau}.} \bibinfo{year}{2011}\natexlab{}.
\newblock \showarticletitle{{Special Section: The third provenance challenge on
  using the open provenance model for interoperability}}.
\newblock \bibinfo{journal}{\emph{Future Gener. Comput. Syst.}}
  \bibinfo{volume}{27}, \bibinfo{number}{6} (\bibinfo{year}{2011}),
  \bibinfo{pages}{737--742}.
\newblock


\bibitem[Soiland-Reyes et~al\mbox{.}(2013)]%
        {Wf4Ever}
\bibfield{author}{\bibinfo{person}{Stian Soiland-Reyes} {et~al\mbox{.}}}
  \bibinfo{year}{2013}\natexlab{}.
\newblock \bibinfo{title}{{Wf4ever Research Object Model}}.
\newblock
\newblock
\urldef\tempurl%
\url{http://wf4ever.github.io/ro/}
\showURL{%
\tempurl}


\bibitem[Stoffers et~al\mbox{.}(2022)]%
        {SMHS22}
\bibfield{author}{\bibinfo{person}{Martin Stoffers} {et~al\mbox{.}}}
  \bibinfo{year}{2022}\natexlab{}.
\newblock \showarticletitle{Integrating Provenance-Awareness into the Space
  Debris Processing System BACARDI}. In \bibinfo{booktitle}{\emph{2022 IEEE
  Aerospace Conference (AERO)}}. \bibinfo{pages}{1--12}.
\newblock
\urldef\tempurl%
\url{https://doi.org/10.1109/AERO53065.2022.9843783}
\showDOI{\tempurl}


\bibitem[{The ILDG Metadata Working Group}(2013)]%
        {qcdml}
\bibfield{author}{\bibinfo{person}{{The ILDG Metadata Working Group}}.}
  \bibinfo{year}{2004--2013}\natexlab{}.
\newblock \bibinfo{title}{{Specification of the QCDml Standard}}.
\newblock
\newblock
\urldef\tempurl%
\url{https://www2.ccs.tsukuba.ac.jp/ILDG/}
\showURL{%
\tempurl}


\bibitem[Wilkinson et~al\mbox{.}(2016)]%
        {FAIR:2016}
\bibfield{author}{\bibinfo{person}{Mark~D. Wilkinson} {et~al\mbox{.}}}
  \bibinfo{year}{2016}\natexlab{}.
\newblock \showarticletitle{{The FAIR Guiding Principles for scientific data
  management and stewardship}}.
\newblock \bibinfo{journal}{\emph{Scientific Data}}  \bibinfo{volume}{3}
  (\bibinfo{year}{2016}), \bibinfo{pages}{160018}.
\newblock


\bibitem[Zhang et~al\mbox{.}(2017)]%
        {zhang_revealing_2017}
\bibfield{author}{\bibinfo{person}{Qian Zhang} {et~al\mbox{.}}}
  \bibinfo{year}{2017}\natexlab{}.
\newblock \showarticletitle{Revealing the detailed lineage of script outputs
  using hybrid provenance}.
\newblock \bibinfo{journal}{\emph{Intl.\ Journal of Digital Curation (IJDC)}}
  \bibinfo{volume}{12}, \bibinfo{number}{2} (\bibinfo{year}{2017}),
  \bibinfo{pages}{390--408}.
\newblock


\bibitem[Zhao et~al\mbox{.}(2006)]%
        {ZWF06}
\bibfield{author}{\bibinfo{person}{Yong Zhao}, \bibinfo{person}{Michael Wilde},
  {and} \bibinfo{person}{Ian~T. Foster}.} \bibinfo{year}{2006}\natexlab{}.
\newblock \showarticletitle{{Applying the Virtual Data Provenance Model}}. In
  \bibinfo{booktitle}{\emph{{IPAW}}} \emph{(\bibinfo{series}{LNCS},
  Vol.~\bibinfo{volume}{4145})}. \bibinfo{publisher}{Springer},
  \bibinfo{pages}{148--161}.
\newblock


\end{thebibliography}

\end{document}